# Access Rate-Dependent Endurance Behavior of Ferroelectric Hafnium Zirconium Oxide Capacitors

*Roberto Guido*, Julie Laguerre*, Pramoda Vishnumurthy, Florian Wunderwald, Kristina Holsgrove, Thomas Mikolajick, Sou-Chi Chang, Alfred Kersch, Uwe Schroeder+*

R. Guido, J. Laguerre, P. Vishnumurthy, F. Wunderwald, T. Mikolajick, U. Schroeder
NaMLab gGmbH, Noethnitzer Strasse 64a, 01187, Dresden, Germany.

K. Holsgrove
School of Mathematics and Physics, Queen's University Belfast, Belfast BT7 1NN, Northern Ireland, United Kingdom

P. Vishnumurthy, T. Mikolajick
Chair of Nanoelectronics, Technical University Dresden, Noethnitzer Strasse 64, 01187 Dresden, Germany

S. C. Chang
Technology Research, Intel Foundry, Hillsboro, OR 97124, USA

A. Kersch
University Applied Sciences, Department of Applied Sciences and Mechatronics, Loth Strasse 34, 80335 München, Germany

*R. Guido and J. Laguerre have equally contributed to this work.

+E-mail: Uwe.Schroeder@namlab.com

**Abstract**

Fluorite-structured ferroelectrics, such as hafnium zirconium oxide ($Hf_{1-x}Zr_xO_2$), are among the most promising materials for developing the next generation of low-power nonvolatile memories. Emulating realistic operating scenarios for ferroelectric memories requires performing access rate-dependent endurance tests. The access rate-dependent reduction of the switched polarization cannot be described as a mere superposition of fatigue and imprint effects, which are observed in conventional endurance and retention tests, respectively. The interplay between these degradation mechanisms and its correlation with the delay time in between the programming pulses remains unclear. Combining electrical testing through voltage pulses with ab initio and drift-diffusion simulations provides a comprehensive perspective on the microscopic origin of the access rate-dependent reduction of the switched polarization of ferroelectric capacitors based on product-relevant 5 nm thick $Hf_{0.5}Zr_{0.5}O_2$ films. The role of oxygen vacancy defects accumulating at the electrode-ferroelectric interface with delay time in reducing the fraction of switchable domains is elucidated considering electron transfer processes which change the defect charge state. Building upon this theoretical understanding, a physics-based model to extrapolate the long-term access rate-dependent endurance performance is developed and experimentally verified for temperatures up to 135°C and delay times up to 2 ms.

## 1. Introduction

The observation of ferroelectric switching in fluorite-structured materials,[1] such as hafnium zirconium oxide ($Hf_{1-x}Zr_xO_2$), renewed the interest in ferroelectrics for developing the next generation of low-power nonvolatile memories.[2–4] Indeed, fluorite-structured ferroelectrics offer significantly better thickness scalability and compatibility with complementary metal-oxide-semiconductor fabrication processes, compared to the classical perovskite-structured ferroelectrics adopted in the first generation of ferroelectric random-access memory (FeRAM) devices.[2–6] Driven by such tremendous application potential, extensive efforts have been devoted through the years to improving the ferroelectric properties of the films, while simultaneously improving the reliability performances and reducing the programming voltages.[4,7] Important aspects on which such optimization efforts focused on are strain optimization through adjustments in the composition and annealing conditions,[8,9] preserving the ferroelectric properties in thinner layers,[10] and engineering the interface with the electrodes.[7,11,12] Year after year, FeRAM arrays with continuously increasing densities and integrating three-dimensional capacitors have been demonstrated.[13–18] Remarkably, the first product-type ready FeRAM device was presented in 2023, promising to meet the requirements of demanding artificial intelligence (AI) workloads.[18]

Despite the proven integration of fluorite-structured thin films product-relevant in FeRAM technologies,[18] reaching full market acceptance requires meeting application-specific reliability performances. Typical reliability testing focuses on the independent assessment of the two key metrics making a nonvolatile ferroelectric memory fully functional: how many times the polarization can be reversed as fast as possible (*i.e.*, endurance), and for how long the macroscopic polarization orientation can be retained before or after switching into the opposite state (*i.e.*, retention), while maintaining the switched polarization above the threshold required for distinguishing the two memory states.[6,11,19,20] Such traditional testing methodologies consider the two extreme cases where a memory cell is accessed continuously within a short delay time, or the data is not accessed for a long time. Hence, despite providing important insights on crucial degradation mechanisms, such as fatigue and imprint,[21,22] standard endurance and retention tests consider these mechanisms as isolated and are, thus, incapable of accounting for the complex interplay between them. Such a limitation is expected to overestimate the memory cell performance in the access rate-specific scenarios relevant for AI workloads, where the cumulative effects of several degradation mechanisms play a crucial role.

In response to the limitation of traditional reliability testing methodologies, an increasing number of recent conference contributions aimed at emulating realistic AI workloads by performing access rate-dependent endurance tests and investigating the reduction of switched polarization of ferroelectric nonvolatile memories based on $Hf_{1-x}Zr_xO_2$ in these scenarios. These investigations were performed considering both single ferroelectric capacitors,[23–26] and their integration into product-type memory arrays.[27] While all these reports converged to the same conclusion that the cumulative effect of imprint and fatigue represents the worst-case scenario, and hence the reduction of switched polarization must be correlated with defect generation and redistribution, the microscopic mechanism behind their interplay remains elusive. The access rate-dependent reduction of switched polarization cannot be described as a mere superposition of fatigue and imprint effects.[23] Ascribing the reduction of switched polarization to domain wall pinning by defects, which are typically identified as oxygen vacancies, and that those are generated upon electric field switching cycling and redistributed during the delay time in between the programming pulses,[23–25,27] is reasonable. However, it remains unclear how the interplay between imprint and fatigue occurs, and how it correlates to with the increasing reduction of switched polarization upon cycling with lower access rates. Understanding how the reduction of switched polarization depends on the delay time in between the programming pulses would provide invaluable insights into long-term access rate-dependent endurance performance, whose modeling is currently limited to a purely empirical description,[27] as well as into identifying possible optimization strategies.

Considering product-relevant ferroelectric capacitors with 5 nm thick $Hf_{0.5}Zr_{0.5}O_2$,[18] this work provides a comprehensive description of the dependency of the endurance behavior on the access rate. The combination of electrical testing through voltage pulses with ab initio and drift-diffusion simulations elucidates the role that charged defects, which are accumulating at the electrode-ferroelectric interface, have on domain nucleation, and therefore, on the increasing reduction of switched polarization upon cycling with lower access rates. Based on the theoretical understanding, a physics-based model is developed to extrapolate the long-term access rate-dependent endurance performance.

## 2. Results and Discussion

The ferroelectric capacitors investigated in this work consist of a 5 nm thick $Hf_{0.5}Zr_{0.5}O_2$ thin film sandwiched between 10 nm thick titanium nitride (TiN) electrodes. The $Hf_{0.5}Zr_{0.5}O_2$ grain orientation aligns with that of the TiN electrodes, as shown in the high-resolution transmission electron microscopy image in **Figure 1**a. A titanium oxynitride ($TiO_xN_{1-x}$) interfacial layer forms at the interface with the bottom electrode mainly due to the TiN oxidation during the atomic layer deposition of the $Hf_{0.5}Zr_{0.5}O_2$ film, and at the interface with the top electrode due to oxygen scavenging from the $Hf_{0.5}Zr_{0.5}O_2$ by this electrode during crystallization annealing.[28,29] The oxygen scavenging by the electrode results in the formation of oxygen vacancies at the electrode-ferroelectric interface.[30] We note that oxygen scavenging might occur from the bottom electrode as well, but the two electrode-ferroelectric interfaces go through a different number of processing steps during the sequential fabrication process of the capacitors. Therefore, the $TiO_xN_{1-x}$ interfacial layers have likely a different stoichiometry. On the other hand, both $TiO_xN_{1-x}$ interfacial layers retain a crystalline structure similar to that of the TiN electrode and are dielectric with a low relative permittivity.[28] Grazing incidence X-ray diffraction performed on the whole capacitor stack after crystallization annealing suggests the absence of monoclinic phase portions within the film (Figure 2b). The Bragg reflection at 30.4° can be attributed to either the ferroelectric orthorhombic phase (space group $Pca2_1$) or the tetragonal phase (space group $P4_2/nmc$). The broadness of the X-ray diffraction peak at this Bragg reflection condition impedes a clear phase portion identification because of the almost coincidence of the main Bragg reflections of the orthorhombic and tetragonal phases.

Moreover, the position of the X-ray diffraction peak can be influenced by strain in the film.[9] The evolution of the ferroelectric hysteresis loop during electric field switching cycling (Figure 1c) is suggestive of the initial coexistence of orthorhombic and tetragonal phases, and the increase of the orthorhombic phase portion upon cycling, which leads the ferroelectric capacitors to achieve a remanent polarization ($P_r$) of 19.3 μC cm$^{-2}$ after $2 \cdot 10^6$ bipolar switching cycles. A switched polarization (*i.e.*, $2P_r$) of ≈50 μC cm$^{-2}$ is expected for a randomly oriented polycrystalline $Hf_{0.5}Zr_{0.5}O_2$ layer.[31] Therefore, it is reasonable to assume that the film portion undergoing ferroelectric switching increases from ≈50% to ≈ 77% upon cycling. Indeed, the transition from a pinched to ferroelectric hysteresis loop upon cycling can be indicative of a progressive tetragonal to orthorhombic phase transformation, but additional effects cannot be excluded,[4,9,32,33] as discussed in the following. We note that the X-ray diffraction peak emerging at 37° is not related to any phase in the $Hf_{0.5}Zr_{0.5}O_2$ thin film, and it can relate to the crystalline $TiO_xN_{1-x}$ interface layer according to powder diffraction file cards from the international center for diffraction.

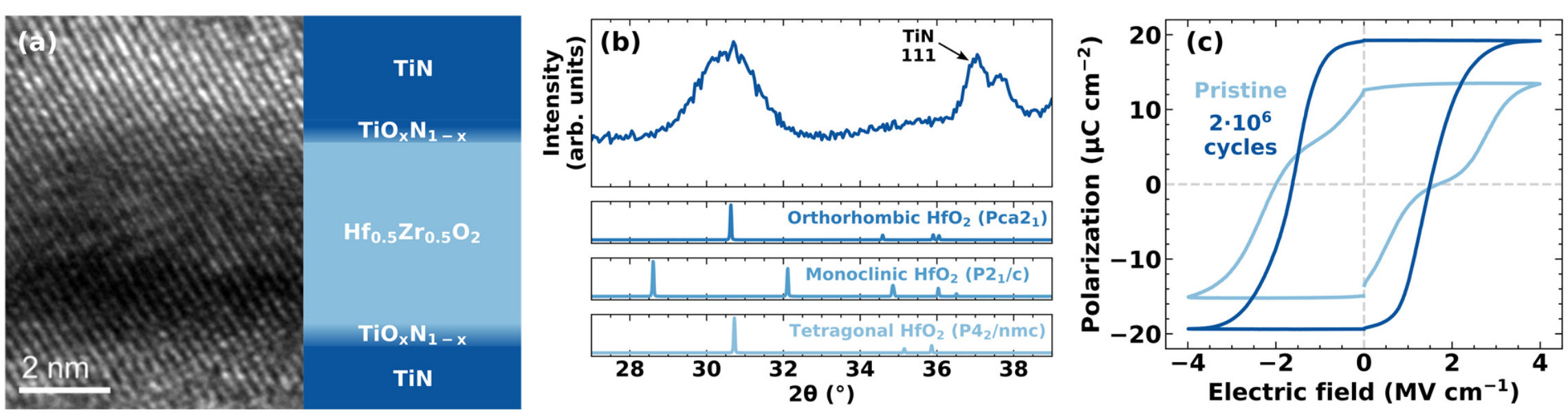

**Figure 1.** (a) Cross-sectional high-resolution transmission electron microscopy of the ferroelectric capacitor stack next to its schematic representation. (b) Grazing incidence X-ray diffraction reported with the reference patterns for the orthorhombic (space group $Pca2_1$), tetragonal (space group $P4_2/nmc$), and monoclinic (space group $P2_1/c$) $HfO_2$ phases. (c) Ferroelectric polarization versus voltage hysteresis loop in the pristine state and after $2 \cdot 10^6$ bipolar switching cycles. The hysteresis loop was extracted after compensating for the leakage and displacement current components through the PUND approach.

Despite the absence of significant $P_r$ increase upon bipolar switching cycling ($\Delta 2P_r$<6 μC cm$^{-2}$), which is commonly referred to as wake-up,[4,33–38] in most of the reports investigating access rate-dependent endurance tests,[23–25] such behavior remains typical for ferroelectric fluorite-structured thin films, especially when combining TiN electrodes with ferroelectric layers thinner than 7 nm.[10] Given that the wake-up phase usually precedes fatigue,[33–35] the influence of the delay time on the endurance performance is analyzed by first focusing on the wake-up behavior, and afterward on the fatigue.

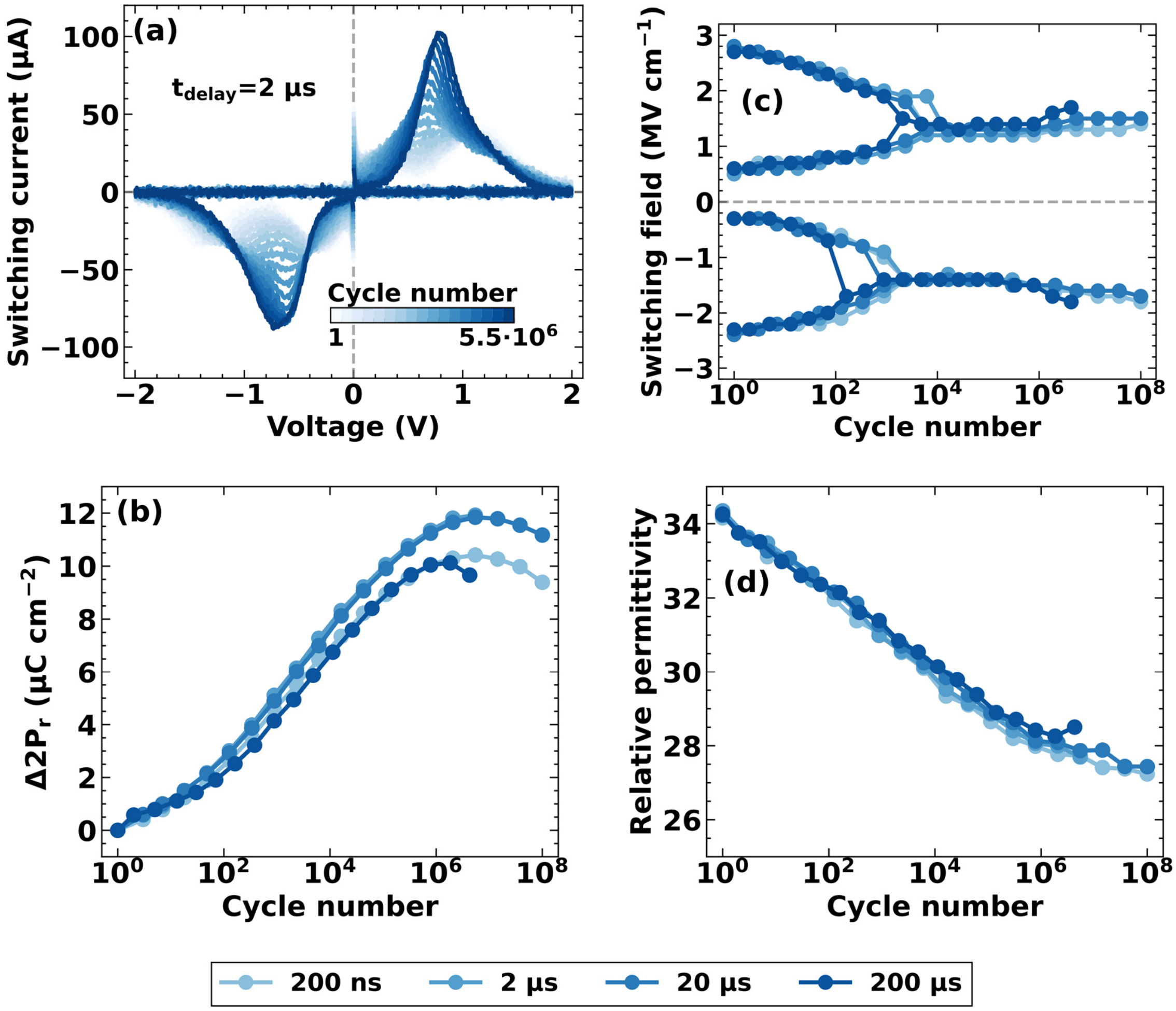

**Figure 2.** (a) Transient switching current versus voltage during wake-up cycling using a delay time ($t_{delay}$) of 2 μs. (b) Relative variation of remanent polarization with respect to the first cycle ($\Delta 2P_r$), (c) switching fields, and (d) relative permittivity versus cycle number for $t_{delay}$ of 200 ns, 2 μs, 20 μs and 200 μs.

Transient switching current measurements performed using positive-up-negative-down (PUND) method,[39] and triangular voltage pulses after an increasing number of bipolar switching cycles demonstrate the initial presence of two clearly distinguishable switching peaks for each voltage polarity, which then progressively merge upon cycling (**Figure 2**a). The progressive coalescence of the switching peaks occurs upon cycling and can be considered completed after ≈$10^4$ cycles independent of the delay time during cycling when this is varied between 200 ns and 200 μs (Figure 2b). We note that a single switching field per voltage polarity, corresponding to the ferroelectric coercive field, is reported in Figure 2b, starting from when the two switching peaks are no longer easily distinguishable from one another. However, the single switching peak corresponds to an asymmetric switching field distribution even after $10^6$ switching cycles. The increase of switchable polarization relative to the first switching cycle ($\Delta 2P_r$) during wake-up also does not show a clear dependence on the delay time and reaches a maximum of 10-12 µC $cm^{-2}$ after $10^6$ cycles in the capacitors investigated (Figure 2c). Hence, the main parameter influencing the wake-up process is the time under which the electric field is applied to the ferroelectric capacitor and not the delay time in between the voltage pulses. Such observation agrees with the possibility of a wake-up process dominated by electric field-induced tetragonal to orthorhombic phase transformations,[4,9,32] which are likely accompanied by strain relaxation. Coherently with this hypothesis and the typically larger relative permittivity of the tetragonal (40) compared to the orthorhombic (28-30) phase,[40] the relative dielectric constant of the capacitors continuously decreases from 34 to 28.5 upon cycling independent of the delay time (Figure 2d). Nonetheless, the contribution to the wake-up process of other electric field-induced phenomena,[33] such as movement of defects and/or charges within the film making easier domain nucleation and/or domain wall motion,[35,38] interface degradation,[41] antipolar to polar phase transitions,[42,43] and ferroelastic switching (*i.e.*, 90° to 180° rotation of domain walls),[36,37] cannot be excluded.

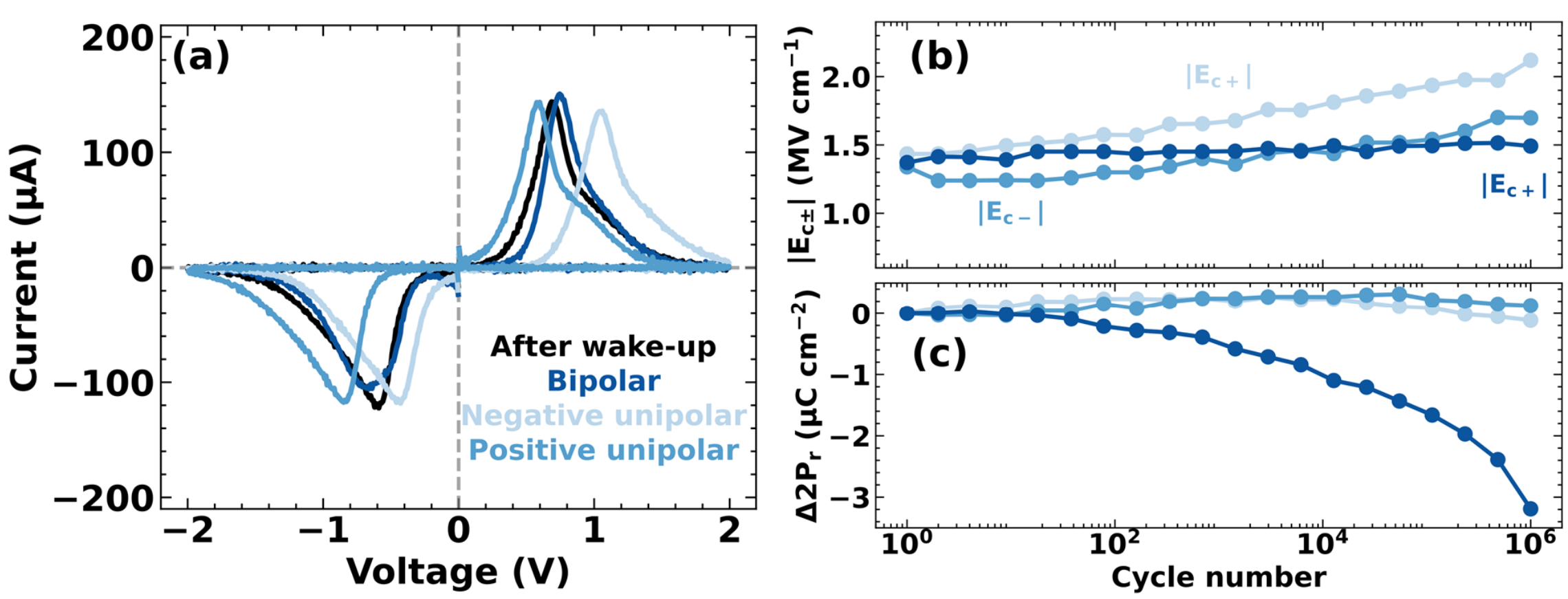

**Figure 3.** (a) Transient switching current versus voltage right after wake-up cycling, and after $10^6$ bipolar switching cycles or unipolar non-switching cycles performed using a positive or negative voltage. (b) Coercive field magnitude ($|E_{c\pm}|$) to switch into the opposite polarization state with respect to the last cycle, and (c) relative variation of remanent polarization with respect to the first cycle ($\Delta 2P_r$) versus cycle number for the three cycling conditions. The color code in (b) and (c) corresponds to that in (a).

In contrast to the wake-up behavior, an increasing reduction of switched polarization with longer delay times between the switching bipolar voltage pulses has been observed for ferroelectric fluorite-structured films undergoing fatigue.[23–27] Some reports ascribe the larger decrease in $\Delta 2P_r$ with lower access-rate to the greater internal electric field originating from the increasing number of charged defects migrating with longer delay times to the electrode-ferroelectric interface.[23–25,27] The direct correlation between $\Delta 2P_r$ and the internal electric field can be tested considering that the latter can be generated by unipolar non-switching cycling. Therefore, after wake-up, three capacitors were subjected to three different cycling tests: bipolar switching cycling, and unipolar non-switching cycling using positive or negative voltage pulses. The transient switching current measurement performed by PUND after $10^6$ cycles show a clear shift of the switching current peaks in the positive(negative) voltage direction with respect to case right after wake-up for the capacitor that underwent negative(positive) unipolar non-switching cycling; whereas, no appreciable shift is observable for the bipolar switching cycling case (**Figure 3**a, and Figure S1). The opposite switching current peak shift with respect to the unipolar non-switching cycling voltage polarity relates to the electrode-ferroelectric interface at which charges accumulate, and hence to the direction of the internal electric field. For instance, applying unipolar non-switching positive voltage pulses to the top electrode results in an accumulation of positive(negative) charges at the top(bottom) electrode-ferroelectric interface due to electron(hole) injection into the top(bottom) electrode, and thus into an internal electric field pointing downwards that increases the electric field magnitude necessary for polarization reversal when applying a negative voltage pulse. Therefore, applying unipolar non-switching cycling increases the coercive field magnitude ($|E_{c\pm}|$) to switch into the opposite polarization state (Figure 3b). The different increase of $|E_{c+}|$ and $|E_{c-}|$ upon cycling for the negative and positive unipolar non-switching cycling likely relates to the asymmetric interfaces of the investigated capacitor stack. We note that these $|E_{c+}|$ and $|E_{c-}|$ would decrease if the unipolar non-switching pulses would move charge defects, such as oxygen vacancies, from one electrode-ferroelectric interface to the other.

Despite no internal electric field being generated upon bipolar switching cycling, this testing condition shows a clear $\Delta 2P_r$ decrease which instead does not occur for both unipolar non-switching cycling conditions given that a sufficiently large voltage is applied to the capacitor to induce full polarization reversal (Figure 3c). Considering that defect generation typically occurs only with ferroelectric switching,[27,44] the $\Delta 2P_r$ decrease upon cycling must be correlated to the increasing defect concentration upon cycling. Nonetheless, a larger imprint would be expected for the bipolar switching test compared to the unipolar non-switching cases according to the macroscopic picture that defects are generated upon electric field switching cycling and redistributed during the delay time in between the programming pulses. The failure of this hypothesis to explain the experimental results in Figure 3 suggests that the interplay between fatigue and imprint in access rate-dependent endurance tests cannot be explained solely by the macroscopic internal bias field related to the superposition of defect generation and migration processes, and must consider the microscopic reasons behind some ferroelectric domains not contributing to the switching process anymore.

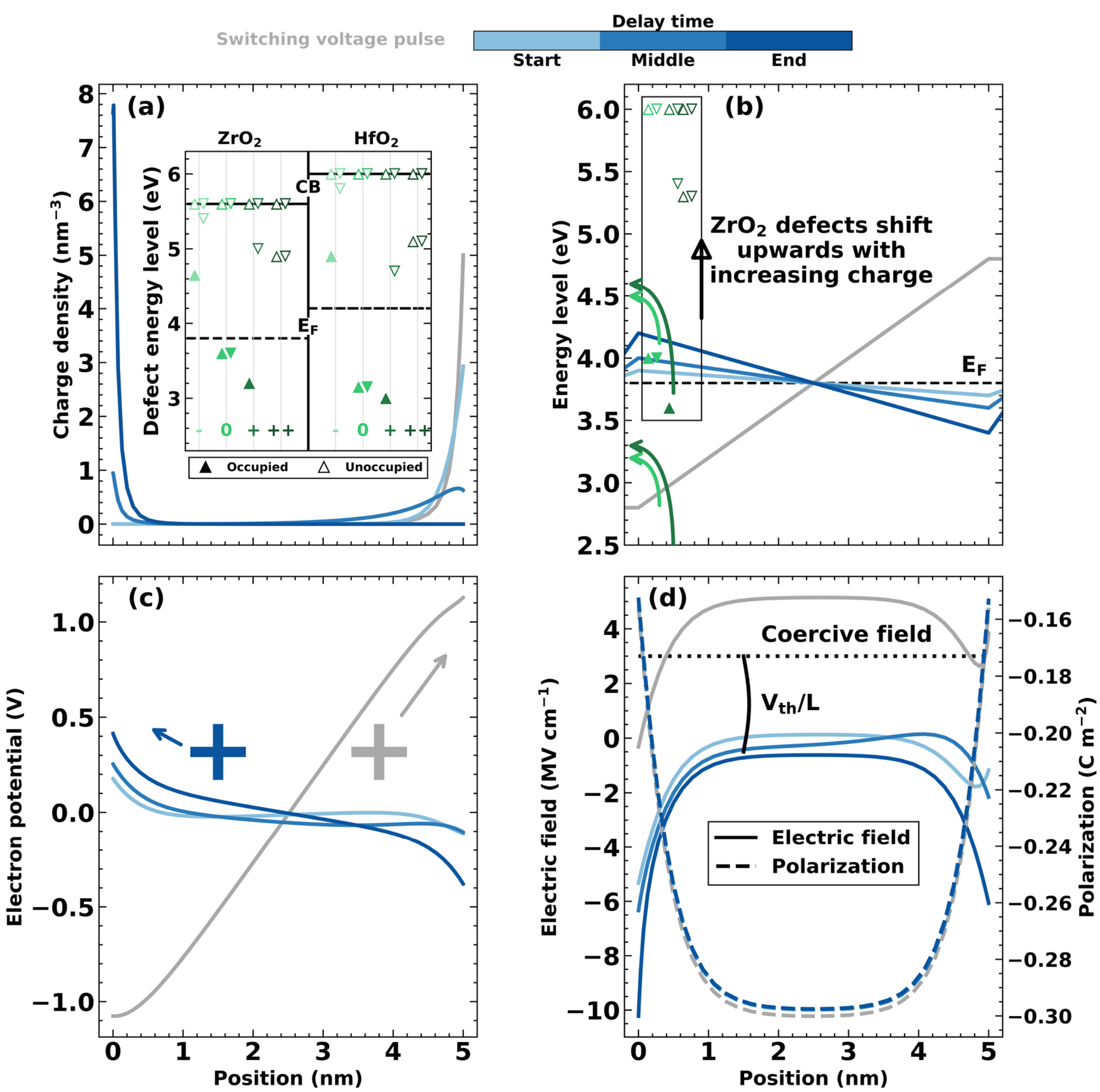

**Figure 4.** (a) Charge distribution during the switching voltage pulse and delay time. The inset shows calculated oxygen vacancy defect levels (triangular symbols). The upward/downward orientation of the triangular symbols refers to the upward/downward spin state of the electrons. The full/empty triangular symbol indicates an occupied/unoccupied level. The level and occupation are shown for possible charge states: singly negative (-), neutral (0), singly positive

(+), and doubly positive (++) charge states. CB and $E_F$ indicate the conduction band minimum and the Fermi level, respectively. (b) Schematic electron potential energy during the switching voltage pulse and delay time, including electrode screening. The applied voltage and subsequent charge redistribution shift defect levels toward the $E_F$, promoting electron emission from oxygen vacancy states. (c) Electrostatic potential during the switching voltage pulse and delay time. The drift direction of positively charged oxygen vacancies is indicated. (d) Polarization profile after polarization reversal (dashed lines) and electric field during the switching voltage pulse and delay time (solid lines), showing an increase in coercive field.

To understand the dependence of the reduction of switched polarization on the delay time during cycling, the sign of the accumulated charge must first be identified. Oxygen vacancy Kohn-Sham defect levels have been calculated from ab initio for monoclinic $HfO_2$ by Broqvist and co-workers,[45,46] for tetragonal $ZrO_2$ using highly accurate Green's function and screened Coulomb interaction (GW) calculations by Jegert et al.[47], and more recently for orthorhombic $HfO_2$ and $ZrO_2$ using Heyd-Scuseria-Ernzerhof screened hybrid functional (HSE06) calculations in direct comparison.[48] Other authors have also reported similar charge-transition levels.[49] The stable charge state depends on the position of the defect levels relative to the Fermi level ($E_F$), as illustrated in the inset of **Figure 4**a. The oxgen vaceny defect levels are located 0.85 eV higher with respect to the conduction band in $ZrO_2$ compared to $HfO_2$. Considering the position of the occupied energy level for the case of a neutral oxygen vacancy, this defect level is closer to the $E_F$ for $ZrO_2$ compared to $HfO_2$. Therefore, it is easier for an oxygen vacancy in $ZrO_2$ to lose an electron compared to $HfO_2$, whose energy levels are much more stable, meaning that electrons need to overcome a higher energy to reach the $E_F$. As discussed in detail below, defects near the electrode are only able to release electrons, an occupation from the electrode in return is very unlikely. Assuming a TiN electrode with a conduction band offset of 2.3 eV,[11] vacancies are expected to occur predominantly in neutral, singly positive, or doubly positive charge states. Indeed, the inset in Figure 4a shows that the occupied defect level of a negatively charged oxygen vacancy is more than 0.5 eV above the $E_F$. Hence, a negatively charged oxygen vacancy loses quickly the negative charge, and therefore the negatively charged state is unstable in the neighborhood of the electrode. Consequently, the experimentally observed charge accumulation is consistent with the formation of positively charged oxygen vacancies resulting from electron transfer from the ferroelectric film to the electrodes.

The effect of positive charge accumulation on the switchable polarization can be understood using the picture proposed by Tagantsev et al.[50] on the depolarization field. A schematic illustration is shown in Figure 4b. In addition, a numerical model was developed that solves coupled boundary-value equations for the polarization and electrostatic potential, combined with drift-diffusion of mobile defects. Representative simulation results are shown in Figure 4a, 4c, and 4d, and additional details of the model can be found in Section S2 in the Supporting Information. Depolarization arising from reduced polarization near the interfaces and incomplete screening in the electrodes generates an electric field inside the film. An additional positive charge further enhances this field. As a consequence, a larger external voltage is required to reach the coercive field, as illustrated in Figure 4d. Therefore, the threshold voltage for polarization reversal increases with the charge accumulation. Domains with local switching thresholds above the applied voltage cannot nucleate anymore; hence, they no longer contribute to switching and give rise to a reduction of switchable polarization. We note that here the focus is on nucleation inhibition rather than domain wall pinning because of the nucleation-limited switching dynamics typically reported for fluorite-structured polycrystalline thin films.[51,52]

Besides the total accumulated charge, its spatial distribution across the film thickness is also important. The simulations shown in Figure 4a and 4c were performed for a fixed amount of accumulated charge of 0.1 C $m^{-2}$ to isolate the effect of charge redistribution during the delay time. During such delay time, positively charged vacancies drift towards the negative pole of the ferroelectric dipole. This redistribution increases the internal electric field, and hence the threshold voltage with delay time.

The redistribution rate depends on the mobility of the charged vacancies. Ab initio molecular-dynamics calculations predict diffusion coefficients in the range of $10^{-10}$-$10^{-12}$ $cm^2$ $s^{-1}$ for oxygen vacancies in $HfO_2$- and $ZrO_2$-based systems.[53] The corresponding drift velocity is of the order of $10^{-5}$-$10^{-4}$ m $s^{-1}$ for a doubly positively charged vacancy at room temperature and an electric field of 1 MV $cm^{-1}$. Therefore, the distance traveled during the experimentally investigated delay times ranges from a fraction of a nanometer to many nanometers. Consequently, vacancies redistribute only weakly at short delay times (*i.e.*, ns- to µs-range). However, they may traverse the entire film thickness at sufficiently long delays (*i.e.*, in the order of seconds). We note that diffusion coefficient predicted by ab initio molecular-dynamics calculations are usually higher than those determined experimentally.[54,55] Such a difference is ascribed to the experimental difficulty of distinguishing the diffusion coefficient of oxygen vacancies with different charge state. Therefore, diffusion coefficients obtained experimentally are an average over the populations of charged and neutral oxygen vacancies. Instead, ab initio molecular-dynamics calculations focus on diffusion of ionized vacancies, which can move significantly faster than neutral defects. However, oxygen vacancy diffusion coefficients depend on the microstructural film features along which these defects can migrate, with grain boundaries and regions near the electrode-ferroelectric interface facilitating the diffusion, and/or the choice of electrode material, which can drive a difference in the defect charge state.[55]

We note that the drift-diffusion model captures the essential physics but remains a simplification. In reality, vacancies can change their charge state and participate in charge transport only while ionized. A kinetic Monte Carlo description including defect charging, discharging, and transport processes would provide a more realistic picture, as demonstrated by Azevedo Antunes et al.[56], but it is computationally prohibitive for the large number of cycles considered here. In such a picture, initially charged vacancies drift first, while neutral vacancies contribute only after becoming ionized. As a result, charge accumulation near the electrode saturates more slowly and extends to longer delay times than predicted by a simple drift-diffusion model.

The origin of the positive charge can be understood from the calculated defect levels. Neutral oxygen vacancies in $ZrO_2$ possess occupied defect levels located approximately 2.0 eV below the conduction band minimum, whereas corresponding levels in $HfO_2$ are located around 2.85 eV below the conduction band minimum. The defect levels in $Hf_{1-x}Zr_xO_2$ span the interval between these limiting cases, depending on the local cation environment. As shown in the inset of Figure 4a, the occupied vacancy levels associated with Zr-rich environments lie close to the electrode Fermi level for a TiN-$Hf_{1-x}Zr_xO_2$ interface. Electrons can therefore be transferred by thermal activation from neutral vacancies to the electrode, converting them into positively charged vacancies. We note that, when a neutral oxygen vacancy releases an electron to the electrode and becomes singly positively charged, the defect level, which is located below the $E_F$ and is filled with two electrons with opposite spin state, turns into one filled level below the $E_F$ and an empty level significantly above the $E_F$. The upward movement of the energy level relates to the structural relaxation occurring in the neighborhood of the singly positive charge state.

A key result of the ab initio calculations is that ionization of a vacancy is accompanied by strong local structural relaxation.[47] The resulting relaxation shifts the empty defect state upward by approximately 1.5 eV. Therefore, electron recapture from the electrode becomes energetically unfavorable, rendering the charge-transfer process effectively irreversible. Vacancies that remain singly occupied can subsequently emit a second electron, leading to the formation of doubly positively charged vacancies. The applied voltage facilitates this process by raising the occupied defect levels relative to the Fermi level. At the same time, the growing positive charge accumulation increases the internal electric field and further raises the occupied defect levels. Consequently, charge generation proceeds not only during the voltage pulse but also during the subsequent delay time.

The accumulated positive charge increases the internal electric field and thereby the external voltage required for polarization reversal. As the threshold voltage rises, an increasing fraction of domains do not nucleate anymore, and hence no longer contribute to switching. The resulting reduction of switchable polarization is hence observed experimentally as reduced polarization after switching. Assuming a logarithmic increase of the threshold voltage with time, and an acceleration factor (β) which accounts for the progressive defect generation upon cycling, the reduction of switched polarization as function of the cycle number (N) can be written as

$$\Delta 2P_r(N) = -\alpha^2 \cdot ln(N^\beta/\tau_0)^2, \quad (1)$$

where α is a proportionality constant, and $\tau_0$ is the ratio between the linear-to-logarithmic charge relaxation crossover time (τ) and the delay time ($t_d$). We note that equation (1) is extracted under the assumption of a logarithmic charge relaxation regime; therefore, it is valid for

$$t_d \cdot N^\beta \gg \tau, \quad (2)$$

and hence the description in equation 1 becomes more accurate with increasing number of bipolar switching cycles. The complete derivation of equation 1 is reported in Section S4 in the Supporting Information. We note that α and β parameters depend on the specific fabrication process and capacitor stack chosen. This work considers only one capacitor stack which was fabricated through an established process; therefore, the experimental access-rate dependent endurance behaviors were fitted through $\tau_0$ (Table S2 and S3 in the Supporting Information) while keeping α and β fixed to $2 \cdot 10^{-9}$ $\mu C^{-2}$ $cm^{-4}$ and 1.7, respectively.

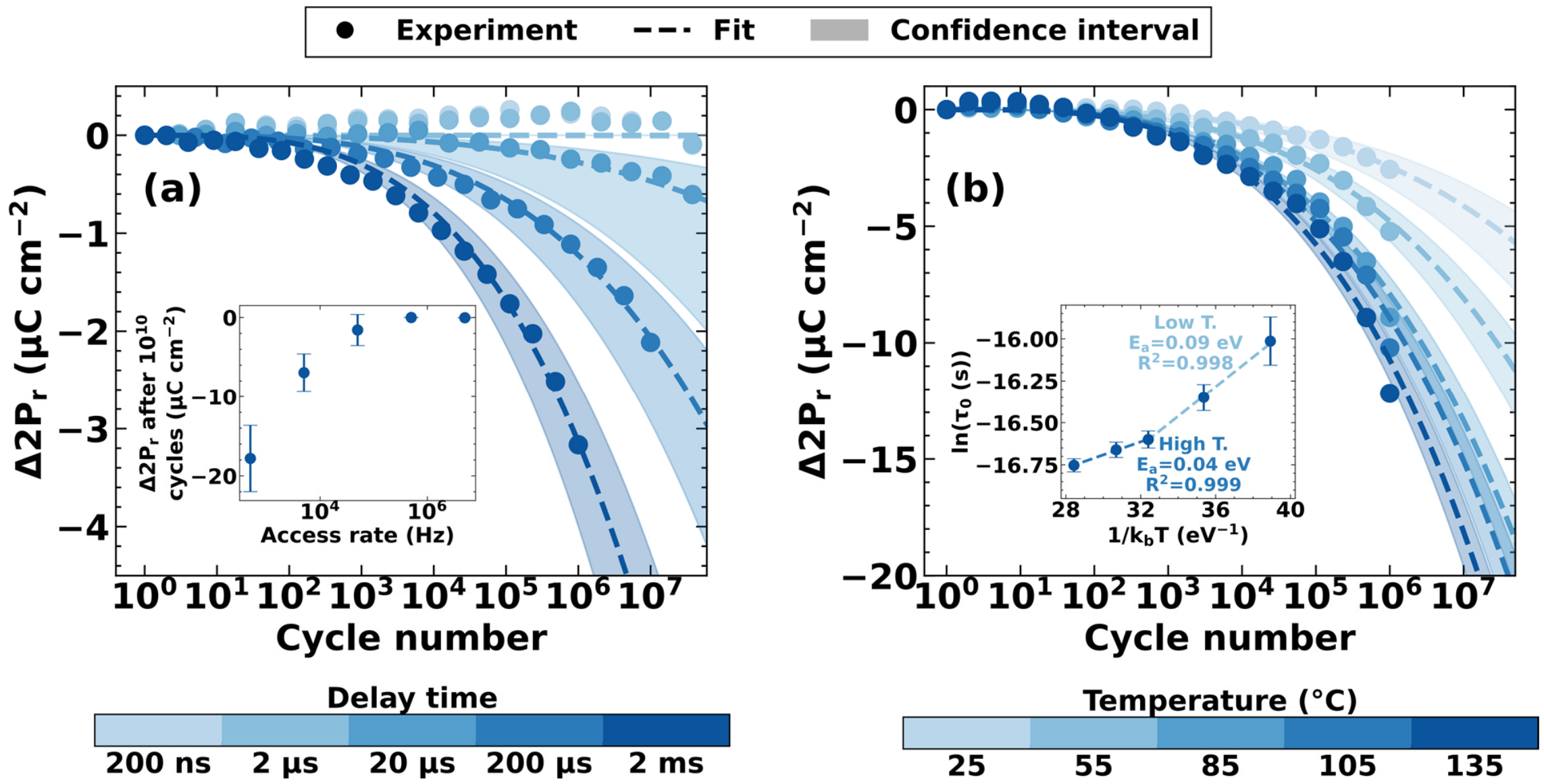

**Figure 5.** Relative variation of remanent polarization with respect to the first cycle ($\Delta 2P_r$) versus cycle number (a) at room temperature and for delay times between 200 ns and 2 ms, and (b) for temperatures between 25 and 135°C and with a delay time of 2 ms. The inset in (a) shows the $\Delta 2P_r$ after $10^{10}$ bipolar switching cycles versus access rate. The inset in (b) shows the natural logarithm of $\tau_0$ in seconds extracted from the fit versus inverse of the temperature multiplied by the Boltzmann constant and its fit in both low and high temperature regions. The fit is shown as dashed line and the confidence interval is shown as a shaded region considering one standard deviation of the fitting parameter.

Equation 1 fairly represents the reduction of switched polarization versus cycle number for all investigated delay times between 200 ns and 2 ms (**Figure 5**a). We note that delay times longer than 2 ms have not been considered since this already exceeds the maximum delay time for embedded dynamic random-access memories.[26] The increasing reduction of switched polarization with lower access rates is obtained by extrapolating to $10^{10}$ bipolar switching cycles (inset in Figure 5a). In contrast with Ettissery et al.[27], the reduction of switched polarization is not expected to decrease for extremely high access rates, and such a phenomenon likely involves additional wake-up effects which are not accounted for by the proposed model because they do not occur in the capacitors investigated in this study. The stability of the capacitors was monitored during the access rate-dependent endurance tests, and both leakage current density and relative permittivity showed no appreciable variation during cycling (Figure S2 in the Supporting Information). In all cases, the relative permittivity stayed almost constant around 27.5, as expected for ferroelectric capacitors embedding films with a major orthorhombic phase content, and indicating no transition from the tetragonal to the orthorhombic phase during this cycling stage.

On the other hand, the validity of the model discussed in this work is supported by the capability to reproduce the reduction of switched polarization during cycling even when performing the endurance test at temperatures up to 135°C (Figure 5b). Extracting the activation energy ($E_a$) for $\tau_0$, two distinct regions can be identified around 85°C. The $E_a$ decreases going from the low to the high temperature region from 0.09 to 0.04 eV. Such low $E_a$ and its temperature-activated reduction suggest that the reduction of switched polarization upon cycling relates to thermal electron de-trapping from defect states rather than to defect migration.[57] Indeed, oxygen vacancy defect migration typically requires energies between 0.5 and 1.2 eV depending on the charge state.[58,59] Hence, the extracted $E_a$ agrees with the picture provided by the ab initio simulations which shows that the main phenomenon governing the reduction of switched polarization is the electron transfer process between oxygen vacancy defect states and the electrode. Given a sufficiently long delay time, defects generated upon cycling can further enhance the reduction of switched polarization when they migrate to the electrode-ferroelectric interface; however, the reduction of switched polarization does not necessarily require such defect migration, since positive charges can accumulate at the electrode-ferroelectric interface simply because of electron transfer processes involving defect states present at this interface. Finally, we note that the $E_a$ extracted in this work is lower than that reported by Ettissery et al.[27] of 0.2 eV. Such discrepancy likely relates to the different mathematical description of the reduction of switched polarization and/or ferroelectric capacitor stack. Nonetheless, both in the work of Ettissery et al.[27] and ours, the $E_a$ is lower than that expected for oxygen vacancy defect migration.

## 3. Conclusion

The access rate-dependent endurance behavior of ferroelectric capacitors has been investigated considering product-relevant 5 nm thick $Hf_{0.5}Zr_{0.5}O_2$ films. The delay time in between the programming pulses applied during cycling does not influence the wake-up process, which is instead dominated by electric field-induced phenomena, such as tetragonal to orthorhombic phase transformation.

On the other hand, the access rate influences the fatigue phenomenon such that a larger reduction of switched polarization is obtained for lower access rates. Comparing the endurance behavior of capacitors undergoing bipolar switching or unipolar non-switching cycling demonstrates that the reduction of switched polarization cannot be explained solely by imprint and actually relates to ferroelectric domains not contributing anymore to the switching process. Through ab initio and drift-diffusion simulations it is proposed that more positive charges accumulate at the electrode-ferroelectric interface with delay time as a consequence of the electron transfer process between oxygen vacancy defect states and the electrode. Such positive charge accumulation increases locally the threshold voltage for domain switching. Therefore, the fraction of domains which do not nucleate anymore increases with delay time, hence resulting in a reduction of switchable polarization. Given a sufficiently long delay time, the defect generation upon cycling further enhances the reduction of switched polarization because these defects migrate to the electrode-ferroelectric interface and take part to electron transfer processes. Based on this theoretical understanding, a physics-based model is developed to extrapolate the long-term access rate-dependent endurance performance and experimentally verified considering temperatures up to 135°C and delay times up to 2 ms.

**4. Methods**

*Ferroelectric capacitor fabrication:* The ferroelectric capacitor stacks were deposited on a p-type silicon substrate. The electrodes were deposited by direct current (DC) sputtering in an ultra-high vacuum sputter cluster tool from Bestec GmbH without additional substrate heating to that resulting from the sputtering process itself. First 30 nm tungsten (W) and a 10 nm TiN layers were deposited to act as bottom electrode. Subsequently, a 5 nm thick $Hf_{0.5}Zr_{0.5}O_2$ layer was deposited by atomic layer deposition at 300°C using an Oxford Instruments OpAL system, with $Zr(C_5H_5)[N(CH_3)_2]_3$ and $Hf(C_5H_5)[N(CH_3)_2]_3$ as the metal-organic precursors with a 1:1 cycle ratio and ozone as the oxidizing agent. Then, a 10 nm TiN top electrode layer was deposited. A rapid thermal annealing (AS-Master 2000 rapid thermal processing tool from Annealsys) was then performed at 500°C in a nitrogen atmosphere to crystallize the $Hf_{0.5}Zr_{0.5}O_2$ film. Following the annealing step, a contact layer of 30 nm W or titanium (10 nm)\platinum (25 nm) was deposited by DC sputtering or electron-beam evaporation (Bestec GmbH tool), respectively. Capacitors with an area of ≈1963 μm$^2$ were finally obtained by patterning the top electrode and contact layers through lithography (Heidelberg Instruments μPG 101 laser lithography tool), inductively coupled plasma etching (Plasmalab System133 from Oxford Instruments), and lift-off in acetone and ultrasonic bath.

*Structural Characterization:* Grazing incidence X-ray diffraction was performed to analyze the crystal structure using a copper (Cu–Kα) X-ray source operated at 40 mA and 40 kV in a D8 Discover diffractometer from Bruker Corporation. A 0.2° slit was placed at the incident and detector sides. The slit at the detector side was a Soller slit, and the incidence angle was fixed to 0.45. Indexing of the Bragg reflections was performed using the powder diffraction file cards 00-038-1420 and 01-084-4872 from the international center for diffraction for TiN and $TiO_xN_{1-x}$, respectively, and the simulated-XRD spectra by Kersch et al.[60] for hafnium oxide crystal phases. Cross-sectional lamellae for transmission electron microscopy (TEM) were prepared using a Tescan-focused ion beam (FIB)-secondary electron microscope Lyra 3. FIB milling was performed with progressively decreasing accelerating voltages of 30, 5, 2, and 1 kV to control the polishing from being initially rough and fast and finally fine and accurate. TEM measurements were acquired on a ThermoFisher Talos F200X.

*Electrical Characterization:* Room temperature electrical measurements were performed using a Cascade Microtech Summit 12000 probe station, whereas high temperature measurements were performed using a Süss MicroTec PA200 probe station. Both probe stations were equipped with a Keithley 4200SCS semiconductor parameter analyzer. Throughout all measurements, the voltage was applied to the top electrode, while the resulting current response was recorded via the bottom electrode. The contact to the bottom electrode was established by electrical breakdown of one capacitor. A different capacitor per endurance test was used. Square pulses with a 5 µs pulse width were applied during each endurance test. Bipolar switching cycling was performed at a voltage of ±2 V for both wake-up and fatigue experiments. Keeping the same voltage amplitude, only one voltage polarity was applied during unipolar non-switching cycling experiments. During unipolar non-switching cycling experiments, the read PUND sequence always started with the opposite voltage polarity compared to that of the cycling pulses. The comparison between bipolar switching and unipolar non-switching cycling experiments was performed fixing the delay time between the cycling pulses to 2 ms, and after $10^6$ wake-up cycles performed at 100 kHz. The access rate-dependent endurance data in Figure 5a,b were acquired after $10^7$ and $10^6$ wake-up cycles performed at 100 kHz, respectively. The relative permittivity upon cycling was extracted from the dielectric current component of the transient current recorded during the PUND measurements which were performed after cycling. In all electrical tests, the PUND pulse sequence was performed at a frequency of 4.6 kHz.

*Simulation:* Details about the ab initio and drift-diffusion simulations are reported in Section S2 in the Supporting Information.

**Acknowledgements**

R.G. was financially supported by the Deutsche Forschungsgemeinschaft (DFG) within the project WUMM2 (project number: 458372836). J.L. was financially supported by LeistungsSchaltEr (project number: 100687609). P.V. was financially supported by Samsung Electronics Co., Ltd. F.W. was financially supported by Intel and DFG within the project D3PO (project number: 505873959). A.K. was also financially supported by DFG within the project D3PO. T.M. and U.S. were financially supported out of the Saxonian State budget approved by the delegates of the Saxon State Parliament. The authors gratefully acknowledge the Gauss Centre for Supercomputing e.V. (http://www.gauss-centre.eu) for funding this project by providing computing time on the GCS Supercomputer SuperMUC-NG at Leibniz Supercomputing Centre (http://www.lrz.de) under Grant No. pn73hi.

Supporting Information

**Access Rate-Dependent Endurance Behavior of Ferroelectric Hafnium Zirconium Oxide Capacitors**

*Roberto Guido*, Julie Laguerre*, Pramoda Vishnumurthy, Florian Wunderwald, Kristina Holsgrove, Thomas Mikolajick, Sou-Chi Chang, Alfred Kersch, Uwe Schroeder+*

R. Guido, J. Laguerre, P. Vishnumurthy, F. Wunderwald, T. Mikolajick, U. Schroeder
NaMLab gGmbH, Noethnitzer Strasse 64a, 01187, Dresden, Germany.

K. Holsgrove
School of Mathematics and Physics, Queen's University Belfast, Belfast BT7 1NN, Northern Ireland, United Kingdom

P. Vishnumurthy, T. Mikolajick
Chair of Nanoelectronics, Technical University Dresden, Noethnitzer Strasse 64, 01187 Dresden, Germany

S. C. Chang
Technology Research, Intel Foundry, Hillsboro, OR 97124, USA

A. Kersch
University Applied Sciences, Department of Applied Sciences and Mechatronics, Loth Strasse 34, 80335 München, Germany

*R. Guido and J. Laguerre have equally contributed to this work.

+E-mail: Uwe.Schroeder@namlab.com

## Section S1: Transient switching current upon bipolar switching or unipolar non-switching cycling

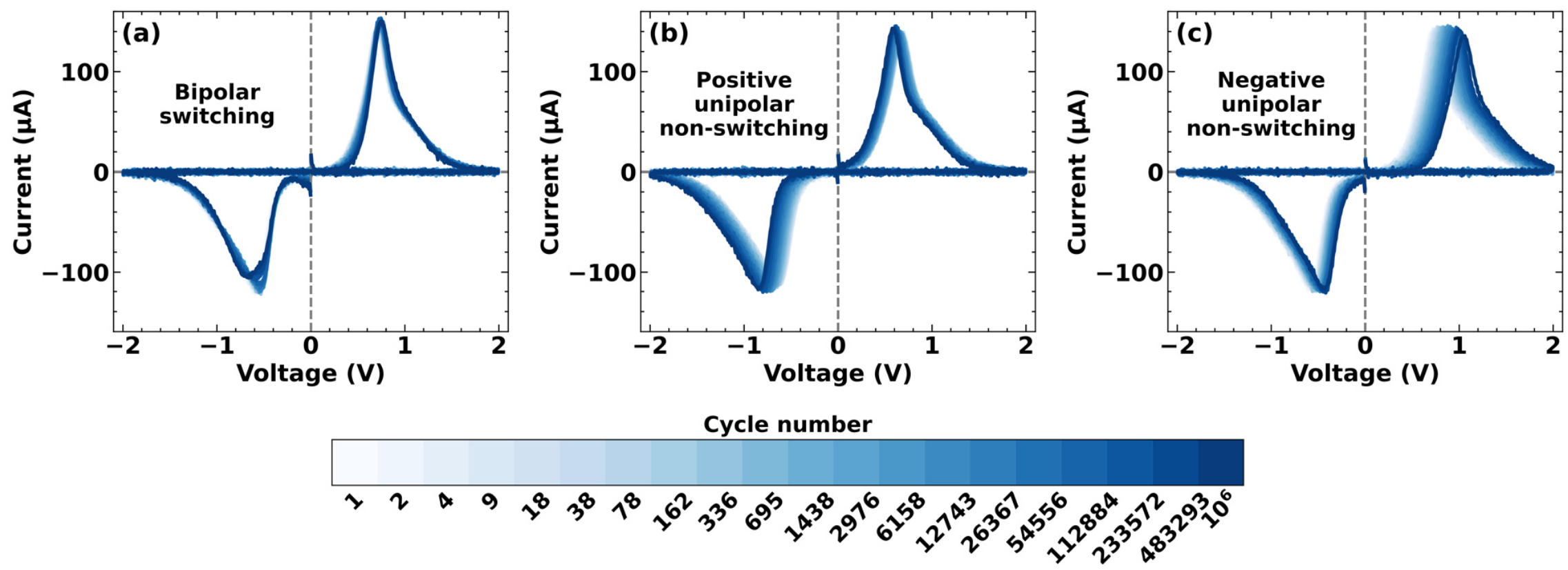

**Figure S1.** Transient switching current upon cycling up to $10^6$ cycles for (a) bipolar switching, (b) positive and (c) negative unipolar non-switching endurance tests.

## Section S2: Details on ab initio and drift-diffusion simulation

### S.2.1 Oxygen vacancy defect level

The oxygen vacancy defect level calculations were carried out spin- polarized using the HSE06 hybrid functional.[1,2]

**Table S1.** Defect level of the spin-up and spin-down states, from valence band to conduction band. HSE06 bandgaps in $ZrO_2$ and $HfO_2$ are 5.6 eV and 6.0 eV. The defect levels in brackets are unoccupied, the other defect levels are occupied.

| $ZrO_2$ | | | | | | $HfO_2$ | | | | | |
|---|---|---|---|---|---|---|---|---|---|---|---|
| $V_O$ | | $V_O^+$ | | $V_O^{++}$ | | $V_O$ | | $V_O^+$ | | $V_O^{++}$ | |
| (5.6) | (5.6) | (5.6) | (5.6) | (5.6) | (5.6) | (6.0) | (6.0) | (6.0) | (6.0) | (6.0) | (6.0) |
| 3.6 | 3.6 | 3.2 | (5.0) | (4.9) | (4.9) | 3.15 | 3.15 | 3.0 | (4.7) | (5.1) | (5.1) |
| 0.0 | 0.0 | 0.0 | 0.0 | 0.0 | 0.0 | 0.0 | 0.0 | 0.0 | 0.0 | 0.0 | 0.0 |

**S.2.2 Ferroelectric electrostatics**

The ferroelectric thin film is described within the one-dimensional Landau-Ginzburg-Devonshire (LGD) framework. The polarization $P(x)$ minimizes the free-energy functional

$$F = \int_0^L \left[\frac{\alpha}{2}P^2 + \frac{\beta}{4}P^4 + \frac{G}{2}\left(\frac{\partial P}{\partial x}\right)^2 - EP\right] dx, \quad \text{(S1)}$$

where $\alpha = -8 \times 10^8$ and $\beta = 9 \times 10^9$ are the Landau coefficients, $G = 2 \times 10^{-10}$ is the polarization-gradient coefficient, and $E = -\partial\phi/\partial x$ is the electric field. Minimization of the free energy yields the Euler-Lagrange equation

$$G\frac{\partial^2 P}{\partial x^2} = \alpha P + \beta P^3 - E. \quad \text{(S2)}$$

The electrostatic potential $\phi(x)$ is determined from Poisson's equation

$$\varepsilon_0\varepsilon_b\frac{\partial^2\phi}{\partial x^2} = \frac{\partial P}{\partial x} - \rho, \quad \text{(S3)}$$

where $\varepsilon_b = 30$ is the background dielectric constant and

$$\rho = zq\,c(x) \quad \text{(S4)}$$

is the density of mobile ionic charge generated by oxygen vacancies with charge state $z = 2$, elementary charge $q$, and concentration $c(x)$. The polarization gradient

$$\rho_b = -\frac{\partial P}{\partial x} \quad \text{(S5)}$$

acts as a bound charge density, producing the depolarization field. This field is partially screened by charge redistribution at the electrodes and by charged defects within the ferroelectric layer.

Two characteristic length scales govern the electrostatic behavior near the interfaces.

The first is the polarization extrapolation length $\lambda$, which determines how rapidly the polarization approaches its bulk value near the electrode interfaces. Physically, it reflects the suppression of the ferroelectric distortion close to the interface. The chosen extrapolation length $\lambda$ is consistent with values found in machine-learned-interaction-potential (MLIP) based molecular dynamics calculations.[3] We choose here a value of $\lambda = 0.5\ nm$.

The second is the electrostatic screening length $\lambda_\phi$, which characterizes the finite screening capability of the electrodes and interfacial regions. In practical systems, this effective length may incorporate intrinsic electrode screening as well as dead-layer effects arising from defective or non-ferroelectric interfacial regions.[4] We choose $\lambda_\phi = 0.5\ nm$.

The polarization boundary conditions are

$$\left.\frac{\partial P}{\partial x}\right|_{x=0} = \frac{P(0)}{\lambda}, \ \left.\frac{\partial P}{\partial x}\right|_{x=L} = -\frac{P(L)}{\lambda}, \tag{S6}$$

while the electrostatic boundary conditions are

$$\left.\frac{\partial \phi}{\partial x}\right|_{x=0} - P(0) + \frac{\phi(0)+\frac{V}{2}}{\lambda_\phi} = 0, \ \left.\frac{\partial \phi}{\partial x}\right|_{x=L} + P(L) - \frac{\phi(L)-V/2}{\lambda_\phi} = 0, \tag{S7}$$

with applied voltage $V$.

**S.2.3 Vacancy drift-diffusion**

To model long-term charge redistribution, positively charged oxygen vacancies are treated as mobile ionic defects. Their concentration evolves according to the continuity equation

$$\frac{\partial c}{\partial t} = -\frac{\partial J}{\partial x}, \tag{S8}$$

where $J$ is the vacancy flux. The flux consists of drift and diffusion contributions,

$$J = \mu c E_{\mathrm{eff}} - D\frac{\partial c}{\partial x}, \tag{S9}$$

with mobility $\mu$ and diffusion coefficient $D$. The latter is related through the Einstein relation

$$D = \frac{\mu k_B T}{zq}. \tag{S10}$$

The vacancies respond not only to the self-consistent electric field but also to additional interaction forces. The effective driving field is therefore written as

$$E_{\mathrm{eff}} = E - E_{\mathrm{img}} + E_{\mathrm{rep}}, \tag{S11}$$

where $E_{\mathrm{img}}$ describes image-force attraction near the electrodes and $E_{\mathrm{rep}}$ accounts phenomenologically for vacancy-vacancy repulsion. The image-force contribution is represented by a screened interaction

$$E_{\mathrm{img}}(x) \propto \sum_n \frac{\exp(-r_n/\lambda_s)}{(r_n+r_0)^2}, \tag{S12}$$

where $r_n$ denotes the distance to successive image charges, $\lambda_s$ is a screening length, and $r_0$ avoids the singularity at the interface. While the attractive force due to the image force is clearly visible in dielectric materials, in ferroelectric materials it is dominated by the electric depolarization field and thus plays a minor role.

To prevent unphysical vacancy accumulation, a weak repulsive contribution is introduced,

$$E_{\mathrm{rep}} = A_{\mathrm{rep}}\frac{\partial}{\partial x}\ln c, \tag{S13}$$

where $A_{\mathrm{rep}}$ is a phenomenological interaction parameter.

No-flux boundary conditions are imposed,

$$J(0) = J(L) = 0, \tag{S14}$$

ensuring conservation of the total vacancy charge

$$Q = \int_0^L c\,(x)\,dx. \tag{S15}$$

**S.2.4 Quasi-static coupling between ferroelectricity and vacancy motion**

The model assumes a clear separation of timescales between polarization relaxation and vacancy migration.

Polarization and electrostatic potential are assumed to relax much faster than ionic defects. Consequently, for a given vacancy distribution $c(x, t)$, the LGD and Poisson equations are solved to equilibrium. The resulting electric field is then used to advance the vacancy distribution during a finite time step.

Thus, the polarization evolves indirectly through the slowly changing vacancy distribution,

$$c(x, t) \rightarrow \rho(x, t) \rightarrow \phi(x, t) \rightarrow E(x, t) \rightarrow P(x, t), \tag{S16}$$

while no explicit polarization switching dynamics is included. The model therefore describes vacancy redistribution, screening evolution, imprint formation, and internal-bias development under voltage cycling, rather than transient ferroelectric switching kinetics.

**S.2.5 Numerical Implementation**

**Spatial discretization:** The one-dimensional film of thickness $L$ is discretized using a non-uniform grid,

$$x_i = \frac{L}{2}[1 - \cos(\pi s_i)], \quad \text{with} \quad s_i = \frac{i}{N_x - 1}. \tag{S17}$$

This cosine mapping provides enhanced resolution near both interfaces where polarization gradients, charge accumulation, and image-force effects are strongest.

**Solution of the LGD-Poisson system:** The coupled LGD and Poisson equations are written as a first-order boundary-value problem and solved using the SciPy solve_bvp solver.

Defining the state vector

$$y = (P, , P_x, \phi\, \phi_x), \tag{S18}$$

the governing equations become

$$\frac{dP}{dx} = P_x,\ \frac{dP_x}{dx} = -\ P + P^3 - \phi_x,\ \frac{d\phi}{dx} = \phi_x,\ \frac{d\phi_x}{dx} = \frac{P_x - \rho}{\varepsilon_b}, \quad \text{(S19)}$$

after nondimensionalization. At each simulation step, the nonlinear boundary-value problem is iteratively solved until convergence.

**Time integration of vacancy transport:** Vacancy motion is computed explicitly in time. For a given electric-field profile obtained from the LGD-Poisson solution, the vacancy flux is evaluated from drift and diffusion contributions. Spatial derivatives are approximated by finite differences, while an upwind scheme is used for the drift term to maintain numerical stability. The continuity equation is discretized as

$$c_i^{n+1} = c_i^n - \Delta t \left(\frac{\partial J}{\partial x}\right)_i. \quad \text{(S20)}$$

Additional numerical stabilization is achieved through:

1. adaptive diffusion enhancement in regions with strong drift,
2. Gaussian smoothing of concentration gradients,
3. concentration positivity constraints,
4. charge renormalization to enforce total-charge conservation.

The resulting vacancy distribution serves as input for the next LGD-Poisson solution.

**Voltage-cycling procedure:** The simulation alternates between pulse and delay periods. During the pulse state,

$$V = V_{\mathrm{amp}}, \quad \text{(S21)}$$

or

$$V = -V_{\mathrm{amp}} \quad \text{(S22)}$$

for bipolar cycling. During the delay period,

$$V = 0. \quad \text{(S23)}$$

For each interval:

1. the LGD-Poisson equations are solved,
2. the electric field is extracted,
3. the vacancy distribution is advanced using drift-diffusion,
4. the updated vacancy charge is fed back into the electrostatic solution.

Repeated cycling therefore produces a gradual redistribution of positively charged oxygen vacancies, resulting in evolving internal fields, asymmetric charge distributions, and cycle-dependent electrostatic bias. Since polarization switching dynamics are not included, all observed temporal evolution originates from defect migration and its electrostatic coupling to the ferroelectric state.

## Section S3: Leakage current density and relative permittivity upon access rate-dependent endurance cycling

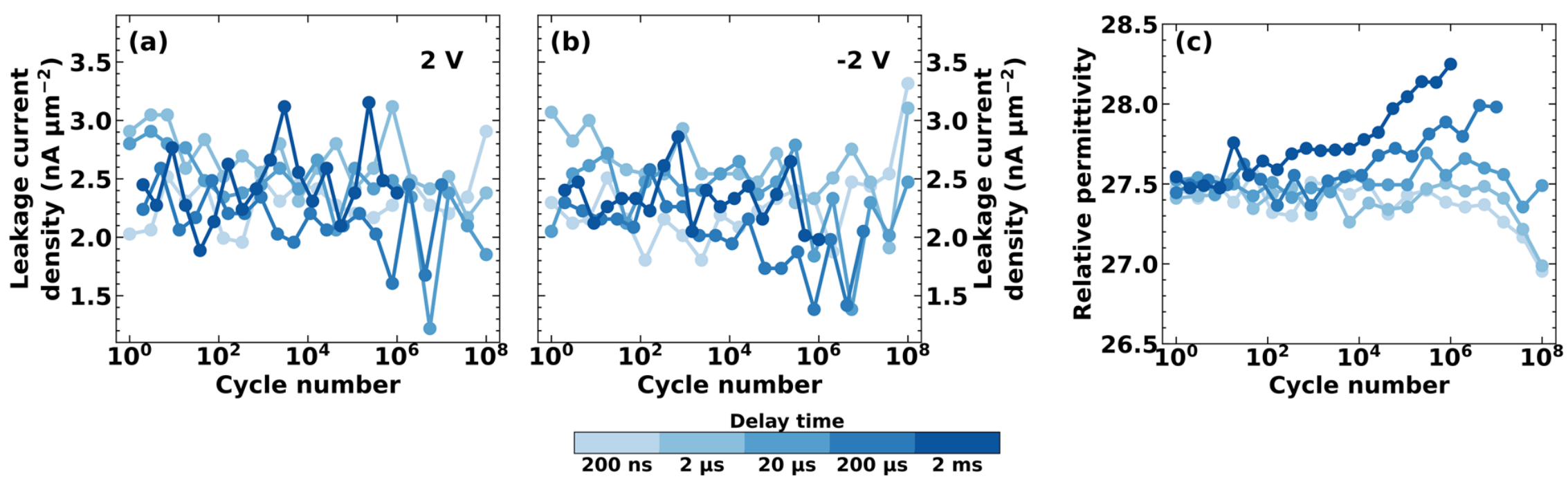

**Figure S2.** Leakage current density extracted at a voltage of (a) 2 or (b) -2 V, and (c) relative permittivity for access rate-dependent endurance cycling performed with delay times between 200 ns and 2 ms after wake-up.

## Section S4: Mathematical derivation of the reduction of switched polarization as function of the cycle number for access rate-dependent endurance tests

A ferroelectric domain switches if locally the applied voltage is larger than the threshold voltage ($V_{th}$), which is directly proportional to the coercive field. The relation between polarization ($P_0$) and the necessary $V_{th}$ to induce polarization reversal of the domain with a given azimuthal angle ($\theta$) can be written as

$$\eta \cdot V_{th}(t) \geq P_0 \cdot \cos\theta(t), \tag{S24}$$

considering $\eta$ as a proportionality constant and that both $V_{th}$ and $\theta$ depend on time (t). Considering the typical nucleation-limited switching dynamics of ferroelectric polycrystalline $Hf_{1-x}Zr_xO_2$ thin films,[5,6] a domain does not switch anymore when its nucleus does not grow beyond the critical size. Therefore, domain nucleation does not occur when the local applied voltage is lower than $V_{th}$. If the $V_{th}$ increases with time, the critical azimuthal angle $\theta$(t) decreases from $\pi/2$ to zero. The azimuthal interval where polarization reversal occurs becomes

$$0 \leq \theta_0(t) \leq \pi/2, \quad \text{(S25)}$$

with $\theta_0(t)$ being the largest azimuthal angle where switching is possible.

Charge accumulation at the interface increases $V_{th}(t)$. It is reasonable to assume a logarithmic dependence of $V_{th}$ on time,[7] which can be written as

$$V_{th}(t) = V_{th,0} \cdot ln(1 + t/\tau), \quad \text{(S26)}$$

Where $V_{th,0}$ is the initial threshold voltage, and $\tau$ is the crossover time between the linear and logarithmic charge relaxation regimes, as Tagantsev et al.[7] proposed in the nearby-electrode injection model. We note that $\tau$ depends on the amount of charge accumulated at the electrode-ferroelectric interface.

Assuming a logarithmic charge relaxation regime,

$$V_{th}(t) = V_{th,0} \cdot ln(t/\tau). \quad \text{(S27)}$$

The increase of $V_{th}(t)$ causes $\theta_0(t)$ to decrease, and hence $\cos(\theta_0(t))$ to increase. Considering equations S24 and S27, we can write

$$\eta \cdot V_{th,0} \cdot ln(t/\tau) = P_0 \cdot \cos\theta_0(t), \quad \text{(S28)}$$

which leads to

$$\cos\theta_0(t) = \frac{\eta \cdot V_{th,0}}{P_0} \cdot ln(t/\tau) = \alpha \cdot ln(t/\tau). \quad \text{(S29)}$$

The switchable polarization of the polycrystalline film is

$$P(t) = \frac{2\pi}{2\pi}\int_0^{\theta_0} P_0 \cdot \cos\theta \cdot \sin\theta \, d\theta = -\int_0^{\theta_0} P_0 \cdot \cos\theta \, d\cos\theta = \int_{\cos\theta_0}^{1} x \cdot P_0 \, dx =$$

$$= P_0 \cdot \frac{x^2}{2}\Big|_{\cos\theta_0}^{1} = P_0 \cdot \frac{1-\cos^2\theta_0}{2}. \quad \text{(S30)}$$

Substituting the time dependency of $\theta_0$ reported in equation S29 into equation S30, the switchable polarization can be written as

$$P(t) = \frac{P_0}{2}(1 - \alpha^2 \cdot ln(t/\tau)^2). \quad \text{(S31)}$$

The time in equation S30 should be the delay time ($t_d$). Considering an endurance test, a delay between the programming pulses is present at every cycle (N). If we consider every cycle as independent, we can neglect the cycling component and just consider the charge accumulation at the electrode-ferroelectric interface due to $t_d$. As discussed in the main manuscript, such independence between cycles does not represent the experimental data, which instead are suggestive of a progressive increase of the charge accumulated at electrode-ferroelectric interface upon cycling. In a simplified case, the effect of the delay time adds up linearly at every cycle, hence:

$$t = t_d \cdot N. \tag{S32}$$

However, this would assume an unperturbed condition upon cycling (*e.g.*, no defect generation upon consecutive polarization reversal), which contradicts experimental observations.[8] Instead, oxygen vacancy defects can generate upon switching cycling and can migrate to the electrode-ferroelectric interface, where more charge tends to accumulate upon cycling and therefore each $t_d$ accumulates more charges than the previous $t_d$. Therefore, considering an acceleration factor (β) and a power law,

$$t = t_d \cdot N^{\beta}. \tag{S33}$$

Hence defining:

$$\tau_0 = \tau / t_d, \tag{S34}$$

equation S31 can be written as

$$P(N) = \frac{P_0}{2}(1 - \alpha^2 \cdot ln(N^{\beta}/\tau_0)^2), \tag{S35}$$

and therefore, the reduction of switched polarization is:

$$\Delta P(N) = -\alpha^2 \cdot ln(N^{\beta}/\tau_0)^2, \tag{S36}$$

With $\tau_0$ that decreases upon increasing $t_d$ such that ΔP(N) increases. The values assumed by $\tau_0$ and their standard deviation for the fits in Figure 5a showing the reduction of switched polarization during endurance tests with different $t_d$ are reported in Table S2. As expected, the standard deviation decreases with $t_d$ because the reduction of switched polarization upon cycling increases and the fit becomes more accurate. Since no reduction of switched polarization is observed for 200 ns and 2 µs, the extracted $\tau_0$ values are purely indicative of a very slow process.

**Table S2.** Values assumed by $\tau_0$ for the fits in Figure 5a for each delay time considered between 200 ns and 2 ms.

| Delay time | $\tau_0$ (ns) | Standard deviation (ns) |
|---|---|---|
| 200 ns | 2385.3 | 49599.2 |
| 2 µs | 613929.1 | 837874415809.9 |
| 20 µs | 330.1 | 139.1 |
| 200 µs | 157.1 | 25.3 |
| 2 ms | 98.0 | 11.2 |

Considering the typical temperature-activation of $\tau$,[7] $\tau_0$ is expected to decrease with temperature. The values assumed by $\tau_0$ and their standard deviation for the fits in Figure 5b showing the reduction of switched polarization during endurance tests performed at different temperature are reported in Table S3.

**Table S3.** Values assumed by $\tau_0$ for the fits in Figure 5b for each temperature considered between 25 and 135°C.

| Temperature (°C) | $\tau_0$ (ns) | Standard deviation (ns) |
|---|---|---|
| 25 | 111.0 | 15.9 |
| 55 | 79.3 | 6.2 |
| 85 | 61.8 | 3.1 |
| 105 | 58.1 | 2.6 |
| 135 | 53.0 | 2.1 |